\documentclass[preprint,showpacs,preprintnumbers,amsmath,amssymb,nofootinbib,showkeys,aps]{revtex4}
\pdfoutput=1
\usepackage{epsfig}

\usepackage{graphicx}
\usepackage{dcolumn}
\usepackage{bm}
\usepackage{amsmath}
\usepackage{amssymb}
\usepackage{latexsym}
\usepackage{color}
\usepackage{hyperref}
\usepackage{mathrsfs}
\usepackage{float}

\newcommand{\be}{\begin{equation}}
\newcommand{\ee}{\end{equation}}
\newcommand{\bea}{\begin{eqnarray}}
\newcommand{\eea}{\end{eqnarray}}

\newcommand{\gQ}{\mathcal{Q}}
\begin{document}


\title{Kinematic Reconstruction of $\Lambda(t)$CDM Models}

\author{P. A. S. Guillen$^{1}$}\email{p.guillen@unesp.br}
\author{J. F. Jesus$^{2,1}$}\email{jf.jesus@unesp.br}
\author{R. Valentim$^{3}$}\email{valentim.rodolfo@unifesp.br}

\affiliation{$^1$Universidade Estadual Paulista (UNESP), Faculdade de Engenharia e Ci\^encias, Departamento de F\'isica - Av. Dr. Ariberto Pereira da Cunha 333, 12516-410, Guaratinguet\'a, SP, Brazil
\\$^2$Universidade Estadual Paulista (UNESP), Instituto de Ciências e Engenharia, Departamento de Ci\^encias e Tecnologia - R. Geraldo Alckmin, 519, 18409-010, Itapeva, SP, Brazil\\
$^3$Universidade Federal de São Paulo (UNIFESP), Departamento de Física, Instituto de Ciências Ambientais, Químicas e Farmacêuticas (ICAQF),\\
Rua São Nicolau, 210, Centro, 17602-496, Diadema/SP , Brazil
}

\def\zt{\mbox{$z_t$}}

\begin{abstract}
In this work, we have \textbf{analysed} two kinematic parametrizations for $\Lambda(t)$CDM models, namely, the linear expansions $\Lambda(z)=\Lambda_0+\Lambda_1z$ and $Q(z)=Q_0+Q_1z$, where $Q$ is the interaction term. In the case of the $Q(z)$ parametrization, we have also tested the particular case of a constant interaction term, $Q(z)=Q_0$. In order to constrain the free parameters of these models, we have used Cosmic Chronometers (CC), SNe Ia data (Pantheon+\&SH0ES) and BAO data. As a general result, we have found weak constraints over the free parameters of the analysed models. In the case of $\Lambda(z)$, we have found for the $\Lambda$ variation parameter, $\Omega_{\Lambda1}\equiv\frac{\Lambda_1}{3H_0^2}=0.02\pm0.14$. In the case of the $Q(z)$ parametrization, we have worked with the dimensionless interaction term $\gQ(z)\equiv\frac{8\pi GQ(z)}{3H_0^3}$, from which we have found $\gQ_0 = 2.2 \pm 2.7$ and $\gQ_1 = -6.2 \pm 7.6$. In the particular case of a constant interaction term, we have found $\gQ_0 = 0.18 \pm 0.7$. All these constraints are at 68\% c.l. The constraints we have obtained are compatible with the standard $\Lambda$CDM model, although still providing a large margin for $\Lambda$ variation.


\end{abstract}
\maketitle



\section{Introduction} 
The $\Lambda$CDM model (comprising a Cold Dark Matter component and a cosmological constant $\Lambda$) has been successful in explaining many observations, including the Cosmic Microwave Background (CMB), Baryon Acoustic Oscillations (BAO), Large Scale Structure (LSS) and SNe Ia magnitudes. With all these observational agreements, $\Lambda$CDM is currently the most widely accepted cosmological framework, being even called as the cosmic concordance model. In this scenario, $\Lambda$ is attributed with driving the acceleration of the Universe, being identified as Dark Energy (DE).

However, $\Lambda$CDM currently is also plagued with many problems. The main problems are the cosmological constant problem (CCP), coincidence problem (CP), that can be seen as theoretical problems. CCP is related to the fact that the quantum vacuum energy contributes to $\rho_\Lambda$ as they share the same equation of state (EOS): $p_{v,\Lambda}=-\rho_{v,\Lambda}$. However, the current estimates on the quantum vacuum energy density given by Quantum Field Theory (QFT) is at least 50 orders of magnitude bigger than the current observational constraints over $\rho_\Lambda$. If both contribute to the same density, they could totally cancel each other. However, it is odd that this cancellation only happens to $10^{50}-1$ parts of $\rho_\Lambda$. This fine-tuning problem is the so-called CCP.

The coincidence problem, however, is more subtle: It refers to the fact that today the matter density and $\Lambda$ density are at the same order of magnitude ($\rho_m\sim\rho_\Lambda$). However, as the matter density dilutes with $a^{-3}$ and $\rho_\Lambda$ is constant, if we look to the last scattering surface, we had $\rho_m\sim10^9\rho_\Lambda$. So, it seems to be a big coincidence that only recently these densities became so similar.

On the other hand, if one allows for $\Lambda$ (or quantum vacuum) to evolve, both these problems could be solved or alleviated if $\Lambda$ was bigger in the past. However, the only way that $\Lambda$ can evolve, while yet assuming total energy conservation, is through interaction with other components, as for instance, dark matter. When the interaction happens between $\Lambda$ and dark matter, these models are called as $\Lambda(t)$CDM models. This idea was proposed quite early in the Modern Cosmology history, it was firstly put forward by Bronstein in 1933 \cite{Bronstein33}. Later, in the 1980s, this idea was recovered as a possible solution to CCP and CP \cite{OzerTaha86,ChenWu90,CarvalhoEtAl92,Abdel-Rahman92,OverdCooper98}. These models were also proposed as a possible solution to the age problem at high redshifts \cite{Jesus06}.   

The $\Lambda$CDM model also faces some problems related to differences in the measured values of certain cosmological parameters, depending on the used method. These differences are called cosmological tensions:  

\begin{itemize}
    \item \textbf{$H_0$ tension}: The Hubble tension refers to the discrepancy between the value of \(H_0\) measured locally ($H_0=73.30 \pm 1.04\, \text{km/s/Mpc}$) from Cepheids and nearby SNe Ia \cite{Riess22}, and the value inferred from the cosmic microwave background (CMB) by Planck satellite, assuming the \(\Lambda\)CDM model (\(H_0=67.4 \pm 0.5\, \text{km/s/Mpc}\)) \cite{Planck:2018vyg}. This difference reaches more than \(5\sigma\). Recently, the Dark Energy Spectroscopic Instrument (DESI) measured a value of \(67.7 \pm 1.5\, \text{km/s/Mpc}\), reinforcing consistency with the CMB but maintaining the tension with local measurements \cite{DESI24}.  

    \item \textbf{$\sigma_8$ tension}: The parameter \(\sigma_8\) describes the amplitude of matter density fluctuations on a scale of \(8 \, h^{-1} \, \text{Mpc}\), quantifying how ``clustered'' matter is on that scale, which corresponds approximately to the typical size of galaxy clusters. The tension arises because the value of \(\sigma_8\) inferred from the CMB within the $\Lambda$CDM model is \(0.812 \pm 0.013\) \cite{dr623}, higher than the value observed at late times from other probes, such as gravitational lensing, which measure lower values, close to \(0.772^{+0.020}_{-0.023}\) \cite{DESI24s8}.  
\end{itemize}

The first $\Lambda(t)$CDM models were initially proposed as a way to solve the CCP and the CP. However, it has been shown that there exists a class of models that can alleviate or solve the $H_0$ and \(\sigma_8\) tensions, although not simultaneously yet. Furthermore, models of interaction between dark matter and dark energy have also been proposed. Generally, these models are defined in terms of an interaction term \(Q(z)\) and an equation of state for dark energy, \(w=\frac{p_x}{\rho_x}\). In this paper, however, we will focus only on the cases where \(w=-1\), i.e., on interaction models between \(\Lambda\) and matter, due to their greater simplicity.  

In this context, the interaction term corresponds:
\begin{equation}
Q(z)=-\frac{\dot{\Lambda}}{8\pi G}
\end{equation}

Once the form of the interaction term is proposed, we can obtain the evolution of the Universe:
\begin{equation}
\dot{\rho}_m+3H\rho_m=-\dot{\rho}_\Lambda=Q(z)
\end{equation}

When formulating such $\Lambda(t)$CDM models, a central question is what is the form of the $\Lambda(t)$ dependence or what is the form of the interaction term. Early vacuum decay models were proposed using dimensional arguments \cite{ChenWu90,CarvalhoEtAl92}, approximations from modified gravity theories, implications from quantum field theory \cite{Rajeev83}, and so on. From these ideas, phenomenological laws for $\Lambda$ were proposed, which could then be compared with observations and guide theoretical physics toward solutions to the aforementioned problems.

By proposing a $\Lambda(t)$CDM model with these arguments, as one is not sure about how the interaction really happens, or even if it happens, one may incur in a bias in the analysis, which can lead to misleading conclusions about the possibility of vacuum decay. To avoid this type of bias, and to obtain indications of vacuum decay in the most direct way possible from the available data, in the present work we will follow a different approach. That is, we propose to reconstruct kinematically \(\Lambda(z)\) and \(Q(z)\), assuming only that these are smooth, linear functions of redshift.

There has been already in the literature some attempts to perform non-parametric reconstructions of the interaction term $Q$ \cite{YangEtAl15,MarttensEtAl20,BonillaEtAl22,EscamillaEtAl23}. These reconstructions, however, focused on general $w(z)$ dependences, which introduce free parameters that involve the analysis. Here we adopt the simpler $\Lambda(t)$CDM scenario and analyse kinematical parametrizations instead of non-parametric ones.

\section{The FLRW Metric}
The Friedmann-Lemaître-Robertson-Walker (FLRW) metric 
arises from the assumptions of the cosmological principle: the Universe is homogeneous and isotropic \cite{weinberg1972gravitation}. The FLRW metric in comoving coordinates $(t,r,\theta,\phi)$ is given by:
\begin{equation}
ds^2 = -c^2 dt^2 + a(t)^2 \left( \frac{dr^2}{1 - k r^2} + r^2 d\Omega^2 \right),
\end{equation}
where $a(t)$ is the scale factor and $k$ is the scalar that represents the spatial curvature.
\subsection{Friedmann Equations}
The Friedmann equations are derived directly by replacing the FLRW metric into Einstein's field equations \cite{peebles1993principles}. For this, it is necessary to assume that the content of the Universe is described as a perfect fluid, so the energy-momentum tensor $T_{\mu\nu}$ takes the form:
\begin{equation}
T_{\mu\nu} = \left(\rho + \frac{p}{c^2}\right) u_{\mu} u_{\nu} + p g_{\mu\nu},
\end{equation}

Replacing the FLRW metric and the energy-momentum tensor into Einstein's equations, we obtain the Friedmann equations, which govern the dynamics of the scale factor $a(t)$. They can be written as:

\begin{equation}
\frac{\dot{a}^{2}}{a^{2}} + \frac{k}{a^{2}} = \frac{8 \pi G}{3} \rho_T
\end{equation}
\begin{equation}
\frac{2 \ddot{a}}{a} + \frac{\dot{a}^{2}}{a^{2}} + \frac{k}{a^{2}} = -8 \pi G p_T
\end{equation}

\subsection{Derivation of the Continuity Equation}

The continuity equation can be derived from the Friedmann equations and the conservation of the energy-momentum tensor, which in General Relativity is expressed as:
\begin{equation}
\nabla_\mu T^{\mu \nu} = 0.
\end{equation}
In the context of the FLRW metric, the temporal component ($\nu=0$) yields the continuity equation:
\begin{equation}
\dot{\rho}_T + 3H (\rho_T + p_T) = 0.
\label{conttot}
\end{equation}
As we are mainly interested in the late evolution of the Universe, we will neglect the radiation contribution. With this assumption, we may write the total energy density $\rho_T$ and total pressure $p_T$ as:
\begin{equation}
\rho_T = \rho_m + \rho_\Lambda, \quad p_T = p_\Lambda = -\rho_\Lambda.
\end{equation}
where $\rho_m=\rho_{dm}+\rho_b$ is the pressureless matter energy density and $\rho_\Lambda$ and $p_\Lambda$ are the density and pressure of the fluid representation of the cosmological constant. Thus, the general continuity equation takes the form:
\begin{equation}
\dot{\rho}_m + 3H \rho_m = -\dot{\rho}_\Lambda.
\end{equation}

As usual, we have included the cosmological constant as part of the energy-momentum tensor, assuming the relation
\begin{equation}
\rho_\Lambda = \frac{\Lambda}{8\pi G} \implies \dot{\rho}_\Lambda = \frac{\dot{\Lambda}}{8\pi G}.
\label{rhoLambda}
\end{equation}

Therefore, the final continuity equation is given by:
\begin{equation}
\dot{\rho}_m + 3H \rho_m = -\dot{\rho}_\Lambda = -\frac{\dot{\Lambda}}{8\pi G}.
\end{equation}
Here, the term $\dot{\rho}_m + 3H\rho_m$ represents the temporal variation of matter density, including dilution due to the expansion of the Universe, while $-\dot{\rho}_\Lambda$ indicates that the variation in vacuum energy density compensates for any loss or gain in matter density. That is, if $\dot{\rho}_\Lambda<0$, the cosmological constant acts as a source for matter energy density. Otherwise, if $\dot{\rho}_\Lambda>0$, $\Lambda$ acts as a sink.

In this context, where interaction between matter and dark energy is allowed, we can define:
\begin{equation}
Q(t) = -\frac{\dot{\Lambda}}{8\pi G}
\end{equation}
where $Q$ is called the interaction term between matter and the cosmological parameter $\Lambda$.

It is common to use the Hubble parameter $H$ to express the Friedmann and continuity equations, with $H = \frac{\dot{a}}{a}$. From now on, we will also neglect the spatial curvature ($k=0$) due to the cosmic microwave background observations indications that the Universe is nearly spatially flat. With these assumptions, the Friedmann equations now read:
\begin{align}
H^{2} &= \frac{8 \pi G}{3}(\rho_m+\rho_\Lambda)\\
2 \dot{H} + 3 H^{2} &= 8\pi G\rho_\Lambda\\
\dot{\rho}_m + 3H \rho_m &= -\dot{\rho}_\Lambda = Q(t) = -\frac{\dot{\Lambda}}{8\pi G}\label{eqcont}
\end{align}

In the present work, aiming to find any evidence for interaction between matter and $\Lambda$, we shall parametrize $\Lambda$ and $Q$ as linear functions of the redshift $z$. In the next section, we show how we obtain the dynamic equations, mainly $H(z)$, from these assumptions, in order to obtain constraints over the interaction.

\section{Parametrizations}
\subsection{\texorpdfstring{$\Lambda(z) = \Lambda_0 + \Lambda_1 z$}{L(z)=L0+L1 z}}
Starting from the linear redshift parametrization, $\Lambda(z) = \Lambda_0 + \Lambda_1 z$, we can express this as the energy density of $\Lambda$ using the relation \eqref{rhoLambda}, giving:
\begin{equation}
\rho_{\Lambda}(z) = \rho_{\Lambda 0} + \rho_{\Lambda 1} z
\label{rhoL1}
\end{equation}
where $\rho_{\Lambda0}\equiv\frac{\Lambda_0}{8\pi G}$ and $\rho_{\Lambda1}\equiv\frac{\Lambda_1}{8\pi G}$. Differentiating $\rho_\Lambda$ with respect to $z$ using the chain rule:
\begin{equation}
\frac{d}{d t} = \frac{d a}{d t} \frac{d z}{d a} \frac{d}{d z}
\label{dadt}
\end{equation}

The scale factor $a$ can be expressed in terms of redshift: $a = \frac{1}{1+z}$, thus, we have: $\frac{d z}{d a} = -(1+z)^2$. Furthermore, since $H = \frac{\dot{a}}{a}$, then $\frac{d a}{d t} = a H$. Replacing these relations into \eqref{dadt}:
\begin{equation}
\frac{d}{d t} = -H (1+z) \frac{d}{d z}
\label{dtdz}
\end{equation}

With this relationship, equation \eqref{eqcont} becomes:
\begin{align}
-H (1+z) \frac{d \rho_m}{d z} + 3 H \rho_m &= Q \label{rhom1cont}\\
\frac{d \rho_{\Lambda}}{d z} &= \frac{Q}{H (1+z)}\label{drhoL1dz}
\end{align}

Replacing (\ref{rhoL1}) into equation (\ref{drhoL1dz}):

\begin{equation}
\frac{d}{d z} \left( \rho_{\Lambda 0} + \rho_{\Lambda 1} z \right) = \frac{Q}{H (1+z)}
\end{equation}
From which we obtain:
\begin{equation}
Q = \rho_{\Lambda 1} H (1+z)
\end{equation}

Once this interaction term is obtained, we can replace it into equation (\ref{rhom1cont}):

\begin{equation}
\frac{d \rho_m}{d z} - \frac{3 \rho_m}{(1+z)} = -\rho_{\Lambda 1}
\end{equation}

Equation above is a first-order inhomogeneous ordinary differential equation (ODE). One way to solve it is by the integrating factor method:

\begin{equation}
\rho_m = c_1(1+z)^{3} + \frac{\rho_{\Lambda 1}}{2} (1+z)
\label{rhom1zc1}
\end{equation}
where $c_1$ is an arbitrary integration constant. In order to obtain this arbitrary constant, an initial condition at $z=0$ is required, which is $\rho_m(0)=\rho_{m0}$, where $\rho_{m0}$ is the current matter density. Thus, we have $\rho_{m0} = c_1 + \frac{\rho_{\Lambda 1}}{2}$ and $c_1=\rho_{m0}-\frac{\rho_{\Lambda 1}}{2}$ and replacing it into Eq. \eqref{rhom1zc1}, we have:
\begin{equation}
\rho_m = \left( \rho_{m0} - \frac{\rho_{\Lambda 1}}{2} \right) (1+z)^{3} + \frac{\rho_{\Lambda 1}}{2} (1+z)
\end{equation}

This furnishes $\rho_m(z)$. Next, we derive the Hubble parameter. For this, we replace $\rho_T$ with $\rho_m$ and $\rho_\Lambda$ into the Friedmann equation expressed in terms of the Hubble parameter. From the $\rho_\Lambda$ parametrization suggested by equation (\ref{rhoL1}), we have $\rho_\Lambda(z) = \rho_{\Lambda 0} - \rho_{\Lambda 1} + \rho_{\Lambda 1}(1+z)$, and the Friedmann equation becomes:
\begin{equation}
H^{2} = \frac{8 \pi G}{3} \left[ \left( \rho_{M 0} - \frac{\rho_{\Lambda 1}}{2} \right) (1+z)^{3} + \frac{3 \rho_{\Lambda 1}}{2} (1+z) + \rho_{\Lambda 0} - \rho_{\Lambda 1} \right]
\label{H21z}
\end{equation}

To work with dimensionless quantities, we divide equation (\ref{H21z}) by $H_0^2$, where $H_0\equiv H(z=0)$ is the Hubble constant.

Evaluating the Friedmann equation today ($z=0$), we find the relation
\begin{equation}
H_{0}^{2} = \frac{8 \pi G}{3}(\rho_{m_0}+\rho_{\Lambda0})=\frac{8 \pi G}{3} \rho_{T0}\equiv\frac{8 \pi G}{3} \rho_{c0}
\label{H02}
\end{equation}
from where we define $\rho_{c0}$ as the current critical density of the Universe. That is, the critical density is the total density of the Universe when the Universe is spatially flat. Finally, dividing (\ref{H21z}) by (\ref{H02}):
\begin{equation}
\left( \frac{H}{H_{0}} \right)^{2} = \frac{1}{\rho_{c0}} \left[ \left( \rho_{m0} - \frac{\rho_{\Lambda 1}}{2} \right) (1+z)^{3} + \frac{3 \rho_{\Lambda 1}}{2} (1+z) + \rho_{\Lambda 0} - \rho_{\Lambda 1} \right]
\label{HH02}
\end{equation}

As usual, we introduce the density parameters in order simplify the above equation:
\begin{align*}
\Omega_{m0} &\equiv \frac{\rho_{m0}}{\rho_{c0}}; \\
\Omega_{\Lambda 0} &\equiv \frac{\rho_{\Lambda 0}}{\rho_{c0}}; \\
\Omega_{\Lambda 1} &\equiv \frac{\rho_{\Lambda 1}}{\rho_{c0}}.
\end{align*}
 With these definitions, equation (\ref{HH02}) becomes:
\begin{equation}
H^{2} = H_{0}^{2} \left[ \left( \Omega_{M 0} - \frac{\Omega_{\Lambda 1}}{2} \right) (1+z)^{3} + \frac{3 \Omega_{\Lambda 1}}{2} (1+z) + \Omega_{\Lambda 0} - \Omega_{\Lambda 1} \right]
\label{H2z}
\end{equation}

Evaluating this equation at $z=0$, we obtain the normalization condition:
\begin{equation}
\Omega_{m0} + \Omega_{\Lambda 0} = 1
\end{equation}

That is, $\Omega_{\Lambda 0} = 1 - \Omega_{m0}$, as usual. Thus, the parameter $\Omega_{\Lambda1}$ does not take part of the normalization condition. We can use this relationship to eliminate the dependence of Eq. (\ref{H2z}) on $\Omega_{\Lambda 0}$. Furthermore, as usual, from now on, we shall drop the subscript `0' for $\Omega_{m0}$. Thus, $H^2$ can now be written as:
\begin{equation}
H^{2} = H_{0}^{2} \left[ \left( \Omega_m - \frac{\Omega_{\Lambda 1}}{2} \right) (1+z)^{3} + \frac{3 \Omega_{\Lambda 1}}{2} (1+z) + 1 - \Omega_m - \Omega_{\Lambda 1} \right]
\end{equation}

That is, since we are working with the spatial flatness assumption, we have as free parameters for this model only the parameters $(H_0,\Omega_m,\Omega_{\Lambda1})$

\subsection{\texorpdfstring{$Q(z) = Q_0 + Q_1 z$}{Q(z) = Q0 + Q1 z}}
As in the previous parametrization, it is useful to continue working with derivatives with respect to $z$. Therefore, the relationship $\frac{d}{d t} = -H(1+z)\frac{d}{d z}$ will remain in the derivations.

Replacing the linear expansion for the interaction term into equation (\ref{eqcont}) and changing to derivatives in $z$, we have:
\begin{align}
\frac{d \rho_{M}}{d z} - \frac{3 \rho_{M}}{(1+z)} &= -\frac{Q_{0} + Q_{1} z}{H (1+z)}\label{contrhoMQ0Q1}\\
\frac{d \rho_{\Lambda}}{d z} &= \frac{Q_{0} + Q_{1} z}{H (1+z)}
\end{align}

The above equations form a system of coupled ODEs involving the functions $\rho_m$, $\rho_\Lambda$, and $H$. However, concerning the tests we aim to perform here, we are mostly interested in obtaining $H(z)$ (or, equivalently, $E(z)$). Therefore, let us rearrange the above equations in order to obtaining an ODE in terms of $H(z)$ alone.

To do this, let us revisit the continuity equation for the total energy density (\ref{conttot}). From this equation, we see that:
\begin{equation}
\frac{d \rho_T}{d t} = -3 H \rho_{M}
\label{contrhototm}
\end{equation}

Equation (\ref{contrhototm}) is expressed with a derivative with respect to $t$, and as in the previous section, it is convenient to use the relationship (\ref{dtdz}) to change the derivative in terms of $z$:
\begin{equation}
\frac{d \rho}{d z} = \frac{3 \rho_{M}}{1+z}
\label{drhomdz}
\end{equation}

Another equation that will aid in simplification is Friedmann equation:
\begin{equation}
H^{2} = \frac{8 \pi G}{3} \rho_T
\label{friedrhot}
\end{equation}

Finally, taking the derivative of Friedmann equation (\ref{friedrhot}):
\begin{equation}
\frac{d H^{2}}{d z} = \frac{8 \pi G}{3} \frac{d \rho_T}{d z}
\end{equation}

Applying the chain rule to the left-hand side of equation above and replacing the derivative $\frac{d \rho_T}{d z}$ on the right-hand side with Eq. (\ref{drhomdz}), we obtain:

\begin{equation}
2 H \frac{d H}{d z} = \frac{8 \pi G \rho_{M}}{1+z}
\label{HdHdz}
\end{equation}

Solving this equation for $\rho_M(z)$:
\begin{equation}
\rho_{M} = \frac{(1+z) H}{4 \pi G} \frac{d H}{d z}
\label{rhoMH}
\end{equation}

We arrive at a relationship between $\rho_M$ and $H$, and from here, we replace the value of $\rho_M$ from (\ref{rhoMH}) into continuity equation (\ref{contrhoMQ0Q1}):

\begin{equation}
\frac{d}{d z} \left( \frac{(1+z) H}{4 \pi G} \frac{d H}{d z} \right) + \frac{3}{(1+z)} \left( \frac{(1+z) H}{4 \pi G} \frac{d H}{d z} \right) = -\frac{Q_{0} + Q_{1} z}{H (1+z)}
\label{eq39}
\end{equation}

Solving the derivative in equation (\ref{eq39}) and multiplying by $\frac{4\pi G}{H(1+z)}$, we obtain:
\begin{equation}
\frac{d^{2} H}{d z^{2}} - \frac{2}{(1+z)} \frac{d H}{d z} + \frac{1}{H} \left( \frac{d H}{d z} \right)^{2} = -\frac{4 \pi G (Q_{0} + Q_{1} z)}{H^{2} (1+z)^{2}}
\label{odeH}
\end{equation}

We arrive at a decoupled ODE. However, it has become a second-order nonlinear ODE. This ODE has no analytical solution, so that, aiming for a numerical solution, it is convenient to divide (\ref{odeH}) by $H_0$ to express it in dimensionless form:
\begin{equation}
\frac{d^2E}{dz^2} - \frac{2}{1+z} \frac{dE}{dz} + \frac{1}{E} \left( \frac{dE}{dz} \right)^2 = -\frac{\mathcal{Q}(z)}{2 E^{2} (1+z)^{2}} = -\frac{\mathcal{Q}_0 + \mathcal{Q}_1 z}{2 E^{2} (1+z)^{2}}
\label{odeE}
\end{equation}
where we define the dimensionless Hubble parameter $E(z) \equiv \frac{H(z)}{H_0}$ and the dimensionless interaction term $\mathcal{Q}(z)$ is defined as:
\begin{equation*}
    \mathcal{Q}(z) \equiv \frac{8\pi G Q(z)}{3H_0^3}
\end{equation*}
Similarly, the following dimensionless parameters are defined:
\begin{align*}
    \mathcal{Q}_0 \equiv \frac{8\pi G Q_0}{3H_0^3} \\
    \mathcal{Q}_1 \equiv \frac{8\pi G Q_1}{3H_0^3}
\end{align*}

In order to numerically solve Eq. \eqref{odeE}, we write a dynamical system that describes the evolution of the dimensionless Hubble parameter $E(z)$ and its derivative $u(z) \equiv \frac{dE}{dz}$ as:
\begin{align}
\frac{dE}{dz}     & =u \label{Ezsys1}\\
 \frac{du}{dz}    & = \frac{u}{1+z} - \frac{u^2}{E} - \frac{\gQ(z)}{2 E^2 (1+z)^2}
 \label{Ezsys2}
\end{align}

The initial conditions correspond to present-day values ($z = 0$):
\be
E(0) = 1, \quad u(0) = \frac{3}{2} \Omega_{m0},
\label{ic}
\ee
where $\Omega_{m0}$ is the current matter density parameter. The initial condition for $E(z)$ is obtained from the definition $E(0)=\frac{H_0}{H_0}=1$, while the initial condition for $\frac{dE}{dz}$ comes from Eq. \eqref{HdHdz}. 

\begin{figure}[ht]
    \centering
    \includegraphics[width=0.49\linewidth]{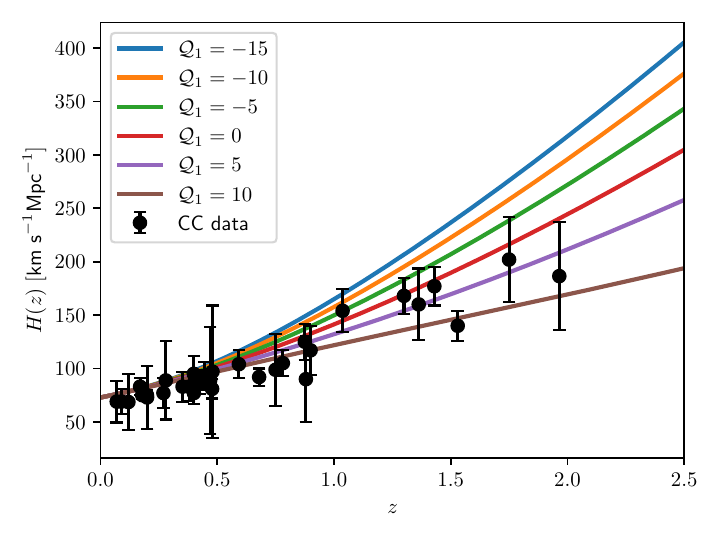}
    \includegraphics[width=0.49\linewidth]{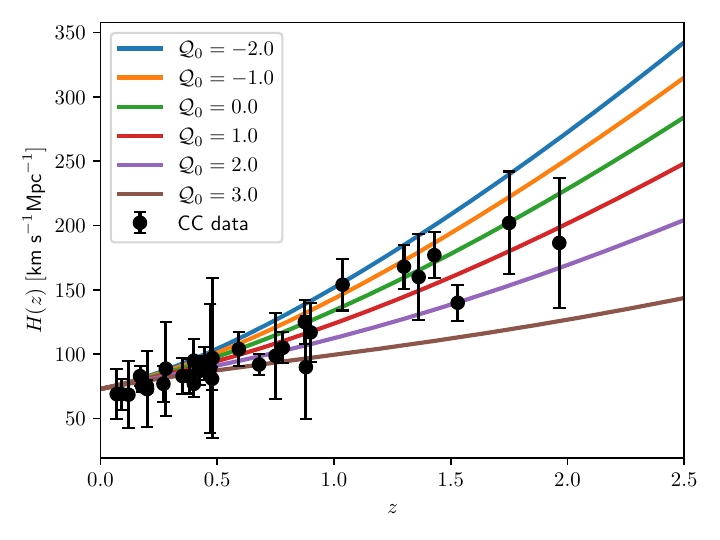}
    \caption{\textbf{Left:} $H(z)$ for $\gQ_0=0$ and varying $\gQ_1$. Other parameters are at their mean values from Tab. \ref{tab:Q0Q1}.  In this case, the curve with $\gQ_1=0$ corresponds to the standard $\Lambda$CDM model. \textbf{Right:} $H(z)$ for $\gQ_1=0$ and varying $\gQ_0$. Other parameters are at their mean values from Tab. \ref{tab:Q0}. In this case, the curve with $\gQ_0=0$ corresponds to the standard $\Lambda$CDM model.}
    \label{fig:Hz}
\end{figure}

As this model has no analytical solution, in Fig. \ref{fig:Hz} we plot some $H(z)$ numerical solutions for varying $\gQ_0$ and $\gQ_1$, while keeping the other parameters at their mean values from Tables \ref{tab:Q0Q1} and \ref{tab:Q0}.

As can be seen from these figures, the larger the values for $\gQ_0$ and $\gQ_1$, the smaller is $H(z)$. Also shown in these figures are the Cosmic Chronometer (CC) $H(z)$ data, described in the next section. As can be noticed, CC data allows for a larger variation of $Q_1$ rather than $Q_0$.

With Eqs. \eqref{Ezsys1}, \eqref{Ezsys2} and initial conditions \eqref{ic}, we may follow with the constraints over the models from observational cosmological data. 

\section{\label{sec: analysis} Analysis and Results}
The observational data used in this study comprise three independent and complementary datasets: (i) the Pantheon+\&SH0ES sample of type Ia supernovae \cite{Pantheon+}, including 1701 light curves from 1550 supernovae in the redshift range $0.001 < z < 2.26$, with SH0ES Cepheid-based calibration; (ii) 32 expansion rate measurements $H(z)$ from cosmic chronometers (CC) \cite{MorescoEtAl22}, based on the relative ages of massive, passively evolving galaxies; and (iii) baryon acoustic oscillation (BAO) data from large-scale surveys such as SDSS, WiggleZ and DES, as compiled by \cite{StaicovaBenisty21}, providing angular and radial distance measurements in the form $d_A(z)/r_d$ over $0.11 < z < 2.4$. As argued by \cite{StaicovaBenisty21}, there is a degeneracy between $H_0$ and $r_d$, so, in order to avoid many assumptions while aiming to constrain them, we have marginalized the BAO posterior distribution function over the combination $\frac{c}{H_0r_d}$. As discussed in \cite{jesus2025high}, these data sets were selected for their statistical robustness and independence from specific cosmological models, allowing for effective tests of $\Lambda$CDM extensions, such as $\Lambda(z)$ variability and dark sector interactions. Their combination offers strong constraints on key cosmological parameters like $H_0$, $\Omega_m$, and the specific free parameters of each model, making them essential for probing deviations from the standard model. Each model was tested using the three data combinations. In particular, the constant interaction case allows us to assess whether current observations favor a persistent energy exchange between dark components or support the standard non-interacting scenario.

Several observational tests can be used to constrain vacuum decay models (for example, \cite{teseJesus}). 
To constrain the models described in the previous section and perform these tests, we use Bayesian statistics, which is based on Bayes’ theorem. This theorem states that the posterior probability density function (PDF) of the free parameters can be written as the product of the prior and the likelihood.

The prior contains all prior knowledge about the parameters — such as physical limits or bounds from other data — while the likelihood is obtained from the dataset being used in the test. Assuming the data uncertainties are normally distributed, the likelihood can generally be expressed as $\mathcal{L}\propto e^{-\chi^2/2}$ \cite{PressEtAl92}, where $\chi^2$ is the frequentist quantity defined as
\begin{equation}
\chi^2=(Y-Y_{obs})C^{-1}(Y-Y_{obs})^T
\end{equation}
with $C$ being the dataset covariance matrix , $Y = Y(\theta_j)$ the model in matrix form with free parameters $\theta_j$, and $Y_{\mathrm{obs}}$ the observational data.

To carry out the Bayesian analyses described above, we employ Markov Chain Monte Carlo (MCMC) sampling, using the \texttt{emcee} package\footnote{\url{https://emcee.readthedocs.io/en/stable/}} \cite{emcee}, following the approach, for instance, in \cite{lima2025new}. \texttt{emcee} implements an affine-invariant ensemble sampler, well-suited for exploring parameter spaces with complex correlations. The likelihood function was built from the total $\chi^2$ across all data sets, incorporating covariance terms where needed. Posterior distributions were obtained by combining the likelihood with uniform priors, from which means, medians, and 68\% (1$\sigma$) and 95\%  (2$\sigma$) confidence intervals were derived. To ensure chain convergence, we adopt the criterium recommended by \cite{emcee}, that is, $n_{steps}>50\tau_{max}$, where $n_{steps}$ is the chain length and $\tau_{max}$ is the maximum autocorrelation time of the free parameters. Then, we discard a burn-in phase of $2\tau_{max}$ and perform a thinning of $\tau_{min}$/2, where $\tau_{min}$ is the minimum autocorrelation time of the parameters, also as suggested by \cite{emcee}.
The use of \texttt{emcee} is particularly advantageous here due to the moderate dimensionality of the parameter spaces and the presence of potential degeneracies, such as between $H_0$ and $Q_0$ or $\Lambda_1$.

\subsection{\texorpdfstring{$\Lambda(z)=\Lambda_0+\Lambda_1z$}{L(z)=L0+L1 z}}
\begin{figure}
    \centering
    \includegraphics[width=.8\textwidth]{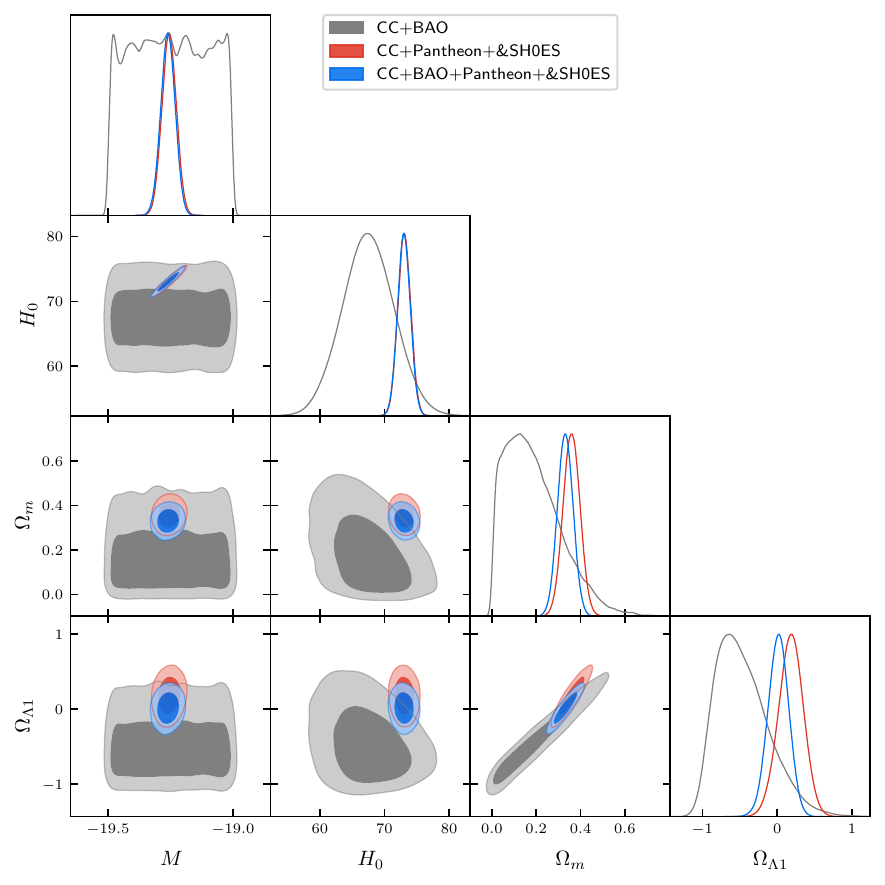}
    \caption{Triangle plot of cosmological parameters for the $\Lambda(z)$ model. Marginal distributions and confidence contours  for the parameters $M$, $H_0$, $\Omega_m$, and $\Omega_{\Lambda1}$ in the model with a time-varying cosmological constant $\Lambda(z) = \Lambda_0 + \Lambda_1 z$. The colors indicate different data combinations: gray (CC+BAO), red (CC+Pantheon+\&SH0ES), and blue (CC+BAO+Pantheon+\&SH0ES).}
    \label{fig:lambdaz_separados}
\end{figure}

\begin{figure}[htbp]
    \centering
    \includegraphics[width=0.8\textwidth]{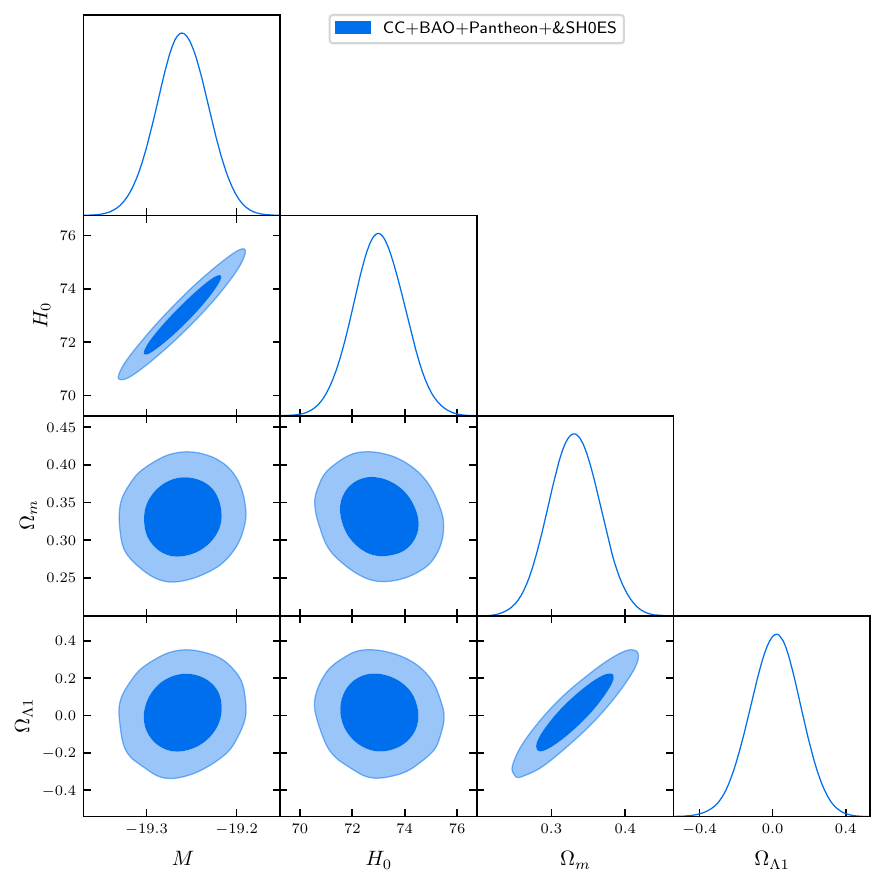}
    \caption{Joint analysis of the cosmological parameters for the $\Lambda(z)$ model. Marginal distributions and confidence contours  for the parameters $M$, $H_0$, $\Omega_m$, and $\Omega_{\Lambda1}$ in the model with a time-varying cosmological constant $\Lambda(z) = \Lambda_0 + \Lambda_1 z$. As in Fig. \ref{fig:lambdaz_separados}, the blue colour indicates the full combination of CC+BAO+Pantheon+\&SH0ES.}
    \label{fig:lambdaz_combinados}
\end{figure}

First of all, let us analyze the model where the cosmological parameter is given as a linear function of the redshift, $\Lambda(z)=\Lambda_0+\Lambda_1z$. The results of this analysis can be seen on Figs. \ref{fig:lambdaz_separados} and \ref{fig:lambdaz_combinados}.

\newpage

As can be seen on Fig. \ref{fig:lambdaz_separados}, for the $\Lambda(z)$ model, the CC + BAO combination (gray contours) results in broad, weakly constrained posteriors, especially for $\Omega_{\Lambda1}$ and $\Omega_m$, which are strongly correlated. This reflects a degeneracy where variations in matter density can be offset by changes in the time-varying cosmological constant. Adding Pantheon+\&SH0ES (red) significantly improves the constraints: contours shrink, particularly for $H_0$ and $\Omega_m$, and $\Omega_{\Lambda1}$ becomes more constrained around zero.

As can be seen on Fig. \ref{fig:lambdaz_combinados}, the full joint analysis (blue) yields the tightest constraints, with the posterior for $\Omega_{\Lambda1}$ nearly symmetric and centered at zero, and a somewhat reduced correlation with $\Omega_m$. The other parameters are almost uncorrelated one with each other. The quantitative results of this analysis can be seen on Tab. \ref{tab:lambda_z}.

\newpage

\begin{table}[htbp]
\centering
\begin{tabular}{cc}
\hline
\textbf{Parameter} & \textbf{68\% and 95\% limits} \\
\hline
$\boldsymbol{M}$                & $-19.260 \pm 0.028 \pm 0.057$ \\
$\boldsymbol{H_0}$ (km/s/Mpc)              & $73.00 \pm 0.98 \pm 2.0$ \\
$\boldsymbol{\Omega_m}$         & $0.331 \pm 0.034 \pm 0.070$ \\
$\boldsymbol{\Omega_{\Lambda1}}$ & $0.02 \pm 0.14 \pm 0.27$ \\
\hline
\end{tabular}
\caption{$\Lambda(z)$ Model – CC+BAO+Pantheon+\&SH0ES}
\label{tab:lambda_z}
\end{table}

As can be seen on Tab. \ref{tab:lambda_z}, at the 1$\sigma$ c.l., this model has $H_0 = 73.00 \pm 0.98$ km/s/Mpc, $\Omega_m = 0.331 \pm 0.034$, and $\Omega_{\Lambda1} = 0.02 \pm 0.14$, with the absolute SNe Ia magnitude $M = -19.260 \pm 0.028$. The large uncertainty in $\Omega_{\Lambda1}$ means that the current data is not enough to discard or confirm a redshift dependence for the cosmological parameter. The small uncertainties in $H_0$ and $\Omega_m$ highlight the constraining power of the full dataset, particularly the influence of Pantheon+\&SH0ES on calibration. The 2$\sigma$ intervals shown in Tab. \ref{tab:lambda_z} are consistent: $H_0 = 73.0\pm2.0$ km/s/Mpc, $\Omega_m = 0.331\pm0.070$, and $\Omega_{\Lambda1} = 0.02\pm0.27$, reinforcing that $\Lambda(z)$ offers no significant improvement over $\Lambda$CDM. The CC+BAO subset alone gives weaker constraints: $H_0 = 67\pm8$ km/s/Mpc, $\Omega_m = 0.19^{+0.25}_{-0.19}$, and $\Omega_{\Lambda1} = -0.45^{+0.73}_{-0.60}$, emphasizing the key role of supernova data.

\subsection{\texorpdfstring{$Q=Q_0+Q_1z$}{Q=Q0+Q1 z}}
\begin{figure}[htbp]
    \centering
    \includegraphics[width=0.8\textwidth]{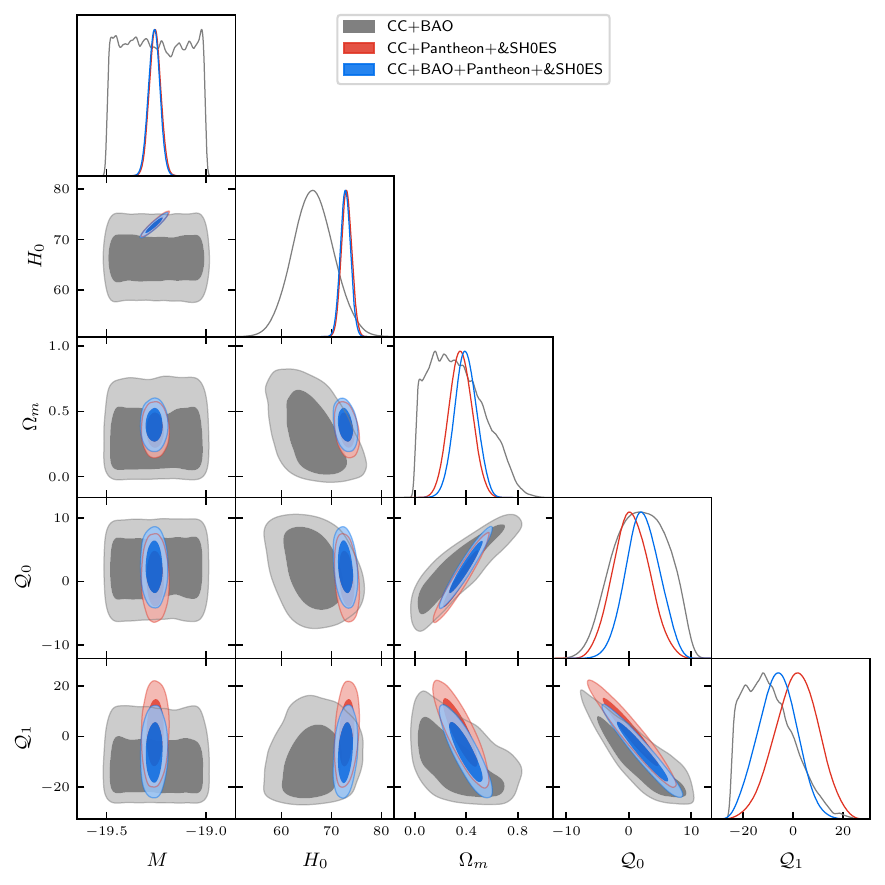}
    \caption{Triangle plot of cosmological parameters for the interacting model $Q(z) = Q_0 + Q_1 z$ for different data combinations. Marginal distributions and confidence contours for the parameters $M$, $H_0$, $\Omega_m$, $\gQ_0$, and $\gQ_1$ in the model with a redshift-dependent interaction between dark matter and dark energy.}
    \label{fig:QOQ1_separados}
\end{figure}

\begin{figure}[htbp]
    \centering
    \includegraphics[width=0.8\textwidth]{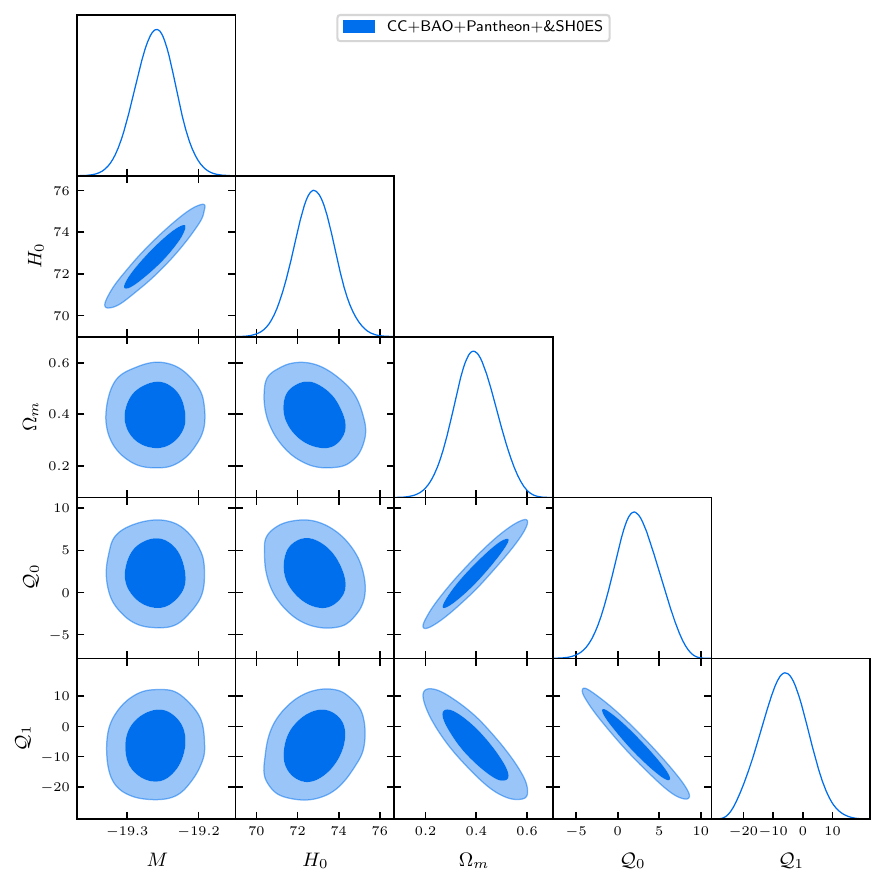}
    \caption{Triangle plot of the joint analysis of the cosmological parameters for the interacting model $Q(z) = Q_0 + Q_1 z$. Marginal distributions and confidence contours for the parameters $M$, $H_0$, $\Omega_m$, $\gQ_0$, and $\gQ_1$ in the model with a redshift-dependent interaction between dark matter and dark energy are shown. As in Fig. \ref{fig:QOQ1_separados}, the blue colour indicates the full combination of CC+BAO+Pantheon+\&SH0ES.}
    \label{fig:QOQ1_combinados}
\end{figure}

The constraints for the model with $Q=Q_0+Q_1z$ can be seen on Figs. \ref{fig:QOQ1_separados} and \ref{fig:QOQ1_combinados}.

\newpage

As can be seen on Fig. \ref{fig:QOQ1_separados}, in the $Q=Q_0 + Q_1z$ model, CC + BAO contours (gray) show even stronger degeneracies than in the $\Lambda(z)$ case. The CC+BAO combination (gray) leads to very broad posterior distributions and pronounced degeneracies among the interaction parameters and $\Omega_m$. In particular, the $(\mathcal{Q}_0,\mathcal{Q}_1)$ and $(\mathcal{Q}_1,\Omega_m)$ planes show highly elongated contours, indicating that the data are unable to independently constrain the redshift dependence of the interaction term.
When Pantheon+\&SH0ES data are added to CC (red contours), the constraint on $H_0$ is improved significantly, reflecting the strong constraining power of
Type Ia supernovae and local distance ladder measurements. However, the interaction parameters remain weakly constrained. In Fig. \ref{fig:QOQ1_combinados} we show the joint CC+BAO+Pantheon+\&SH0ES analysis.

As can be seen on Fig. \ref{fig:QOQ1_combinados}, the full data combination (CC + BAO + Pantheon+\&SH0ES) lead to narrower posteriors, though strong degeneracies persist, mainly in the planes $\gQ_0-\gQ_1$, $\gQ_0-\Omega_m$ and $\gQ_1-\Omega_m$. Let us see the full quantitative results in Tab. \ref{tab:Q0Q1}.

\newpage

\begin{table}[htbp]
\centering
\begin{tabular}{cc}
\hline
\textbf{Parameter} & \textbf{1$\sigma$ and 2$\sigma$ limits} \\
$\boldsymbol{M}$ & $-19.261 \pm 0.028 \pm 0.057$ \\
$\boldsymbol{H_0}$ (km/s/Mpc) & $72.8 \pm 1.0 \pm 2.0$ \\
$\boldsymbol{\Omega_m}$ & $0.395 \pm 0.084 \pm 0.17$ \\
$\boldsymbol{\mathcal{Q}_0}$ & $2.2 \pm 2.7 \pm 5.3$ \\
$\boldsymbol{\mathcal{Q}_1}$ & $-6.2 \pm 7.6 \pm 15$ \\
\hline
\end{tabular}
\caption{$Q = Q_0 + Q_1 z$ Model – CC+BAO+Pantheon+\&SH0ES}
\label{tab:Q0Q1}
\end{table}

As can be seen in Tab. \ref{tab:Q0Q1}, at 1$\sigma$ the full dataset yields
$H_0 = 72.8 \pm 1.0$ km/s/Mpc and $\Omega_m = 0.395 \pm 0.084$,
consistent with the values inferred from supernova-dominated analyses as, for instance, \cite{Pantheon+}. The interaction parameters present central values slightly displaced from zero,
namely $\mathcal{Q}_0 = 2.2 \pm 2.7$ and $\mathcal{Q}_1 = -6.2 \pm 7.6$, but with
large uncertainties that render both parameters statistically compatible with
no interaction.

At the $2\sigma$ confidence level, the allowed ranges widen considerably,
$\mathcal{Q}_0 = 2.2 \pm 5.3$ and $\mathcal{Q}_1 = -6.2 \pm 15$, reinforcing the
conclusion that current data do not provide significant evidence for a
redshift-dependent coupling in the dark sector or even for a coupling in the dark sector, in the context of this model.


Aiming to improve the constraints over the interaction term in this model, we will now analyse a particular case, where we assume $Q_1=0$. That is, a model with constant interaction term.

\subsection{\texorpdfstring{$Q=Q_0$}{Q=Q0}}
\begin{figure}[htbp]
    \centering
    \includegraphics[width=0.8\textwidth]{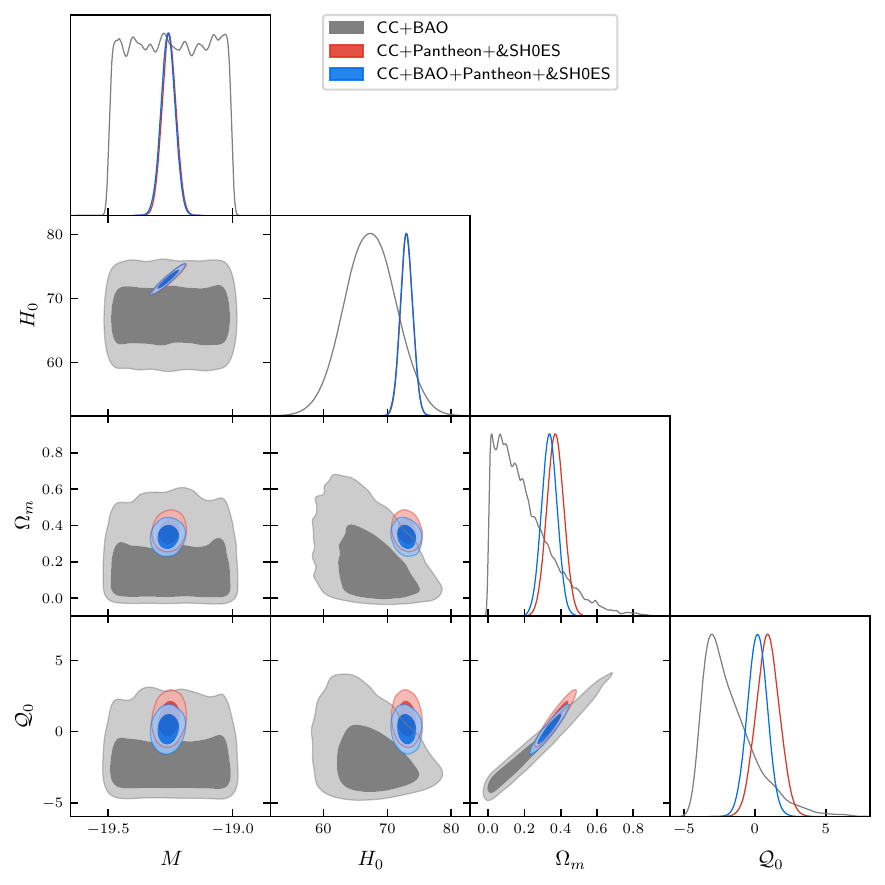}
    \caption{Triangle plot of different data combinations for the interacting model with constant $Q_0$. Marginal distributions and confidence contours for the parameters $M$, $H_0$, $\Omega_m$, and $\gQ_0$ in the model with a constant interaction between dark matter and dark energy.}
    \label{fig:QO_separados}
\end{figure}

\begin{figure}[htbp]
    \centering
    \includegraphics[width=0.8\textwidth]{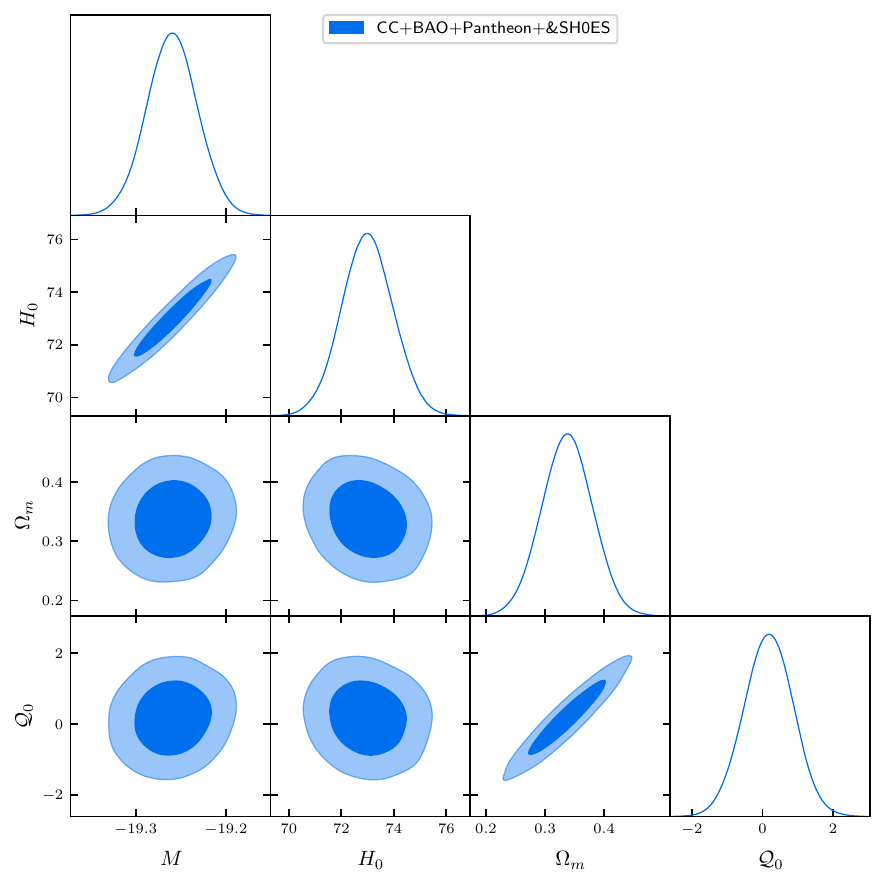}
    \caption{Triangle plot of cosmological parameters for the interacting model with constant $\gQ_0$, in the case of CC+BAO+Pantheon+\&SH0ES data joint analysis. Marginal distributions and confidence contours for the parameters $M$, $H_0$, $\Omega_m$, and $\gQ_0$ in the model with a constant interaction between dark matter and dark energy.}
    \label{fig:QO_combinados}
\end{figure}

Let us now analyse a model where $Q_1=0$, that is, $Q(z)=Q_0$. The results of this analysis can be seen on Figs. \ref{fig:QO_separados} and \ref{fig:QO_combinados}.

\newpage

As shown in Fig. \ref{fig:QO_separados}, fixing $Q_1=0$ leads to a clear improvement in the statistical constraints of the model when compared to the redshift-dependent interaction scenario. For the CC+BAO combination (gray
contours), the posteriors remain relatively broad, but the degeneracies observed
previously are significantly reduced, particularly in the
$(\mathcal{Q}_0,\Omega_m)$ plane. This indicates that a constant interaction term
is more easily accommodated by low-redshift expansion data.

The inclusion of Pantheon+\&SH0ES data (red contours) dramatically sharpens the constraints. In this case, the posterior distribution for $H_0$ becomes well localized, and the interaction parameter $\mathcal{Q}_0$ exhibits a nearly Gaussian distribution centred close to zero. This highlights the crucial role played by supernova data and local distance ladder measurements in breaking parameter degeneracies.

As can be seen in Fig. \ref{fig:QO_combinados}, the contours obtained from the full
data combination are compact and symmetric, indicating a well-constrained
parameter space. The interaction parameter $\mathcal{Q}_0$ shows a posterior
distribution strongly peaked around zero and exhibits only mild correlation
with $\Omega_m$, closely resembling the behaviour found in the $\Lambda$CDM
scenario. This suggests that, within current observational precision, a
constant interaction does not significantly alter the background cosmological
evolution.

\begin{table}[htbp]
\centering
\begin{tabular}{cc}
\hline
\textbf{Parameter} & \textbf{1$\sigma$ and 2$\sigma$ limits} \\
$\boldsymbol{M}$ & $-19.260 \pm 0.028 \pm 0.057$ \\
$\boldsymbol{H_0}$ (km/s/Mpc) & $73.00 \pm 0.98 \pm 2.0$ \\
$\boldsymbol{\Omega_m}$ & $0.338 \pm 0.043 \pm 0.086$ \\
$\boldsymbol{\mathcal{Q}_0}$ & $0.18 \pm 0.7 \pm 1.4$ \\
\hline
\end{tabular}
\caption{$Q = Q_0$ Model – CC+BAO+Pantheon+\&SH0ES}
\label{tab:Q0}
\end{table}

As reported in Tab.~\ref{tab:Q0}, the full dataset yields
$H_0 = 73.00 \pm 0.98$ km/s/Mpc and $\Omega_m = 0.338 \pm 0.043$ at $1\sigma$,
while the interaction parameter is constrained to
$\mathcal{Q}_0 = 0.18 \pm 0.70$. The central value of $\mathcal{Q}_0$ is fully
consistent with zero, and the uncertainty range indicates no statistically
significant detection of interaction. At the $2\sigma$ level, the allowed
interval $\mathcal{Q}_0 = 0.18 \pm 1.4$ further reinforces this conclusion.

\section{\label{sec: conclusion}Conclusion}
In this work, we performed a Bayesian analysis of alternative models to $\Lambda$CDM in order to investigate a possible variation of the cosmological constant, due to an interaction in the dark sector of the universe. One model we have proposed to investigate this possible variation is $\Lambda(z) = \Lambda_0 + \Lambda_1 z$. Other way that we have investigated this possible variation was through a linear interaction term $Q(z) = Q_0 + Q_1 z$ and its constant version $Q = Q_0$. The analysis was carried out through Bayesian inference and MCMC sampling of three independent and complementary observational datasets: Type Ia Supernovae (Pantheon+\&SH0ES), Cosmic Chronometers (CC), and Baryon Acoustic Oscillations (BAO).

The results have shown that the combined dataset (CC+BAO+Pantheon+\&SH0ES) imposes the strongest constraints on the model parameters. For the model with a variable cosmological constant, $\Lambda(z)=\Lambda_0+\Lambda_1z$, the results indicate that the parameters $H_0$ and $\Omega_m$ are well determined, with $H_0 = 73.0 \pm 1.0$ km/s/Mpc and $\Omega_m = 0.331 \pm 0.034$, while the additional parameter $\Omega_{\Lambda1}$ remains weakly constrained. 

In the scenario of a general interaction term between the dark sectors with linear redshift dependence, $\gQ(z)=\gQ_0+\gQ_1 z$, the interaction parameters proved to be weakly constrained, with $\gQ_0 = 2.2 \pm 2.7$ and $\gQ_1 = -6.2 \pm 7.6$ at 1$\sigma$ c.l., and there were strong degeneracies between $\gQ_0$, $\gQ_1$, and $\Omega_m$.  Although the standard cosmological parameters remain reasonably well determined, with $H_0 = 72.8 \pm 1.0$ km/s/Mpc and $\Omega_m = 0.395 \pm 0.084$, the interaction sector remains effectively unconstrained. This behaviour indicates that current background data are not sufficient to simultaneously determine both components of a redshift-dependent interaction.

In contrast, the constant interaction model, $\gQ(z)=\gQ_0$, yields the most restrictive and stable constraints among these interacting scenarios. For the full data combination, the interaction parameter is found to be $\gQ_0 = 0.18 \pm 0.7$, centred at zero and exhibiting low correlation with the remaining cosmological parameters, while the standard parameters are tightly constrained at $H_0 = 73.00 \pm 0.98$ km/s/Mpc and $\Omega_m = 0.338 \pm 0.043$. These results reinforce the compatibility of the current observational data with the $\Lambda$CDM model, although they are not sufficient to exclude the possibility of a nonzero interaction in the dark sector.

In order to further constrain the interaction parameters and reduce the remaining degeneracies, future work may include cosmic microwave background data, as provided by the Planck collaboration \cite{Planck:2018vyg}. In addition, there is the possibility of extending this analysis with non-parametric reconstruction techniques such as Gaussian Processes \cite{Seikel:2012uu}, allowing for a more model-independent investigation of dark sector interactions. All these possibilities indicate that model-independent and hybrid approaches to dark sector phenomenology constitute a promising avenue for future research.

\section*{Appendix}
From the equation for the Hubble parameter $H(z)$:
\begin{equation}
H''(z) - \frac{2H'(z)}{1+z} + \frac{[H'(z)]^2}{H(z)} = -\frac{4\pi G Q(z)}{(1+z)^2 H(z)^2}
\label{eqHinter}
\end{equation}

Let us assume that the interaction term can be negligible. In this case, we recover the equation for the standard $\Lambda$CDM model without energy exchange ($Q_0 = 0$), that is:
\begin{equation}
H''(z) - \frac{2H'(z)}{1+z} + \frac{[H'(z)]^2}{H(z)} = 0
\end{equation}

This nonlinear ODE can be solved with the substitution $y = H^2$. In this case, it can be written as:
\begin{equation}
y''(z) - \frac{2y'(z)}{1+z} = 0
\end{equation}
This linear equation can be easily solved, as there is no $y$ term, only $y$ derivatives. The solution to this equation, after replacing $y=H^2$ is:
\begin{equation}
y(z) = H^2= H_0^2 [\Omega_\Lambda + \Omega_m (1+z)^3]
\end{equation}
This solution inspires us to try the same substitution $y=H^2$ on Eq. \eqref{eqHinter}. With this substitution, we find:
\begin{equation}
y''(z) - \frac{2y'(z)}{1+z} = -\frac{8\pi G Q(z)}{(1+z)^2 \sqrt{y(z)}}
\end{equation}
In this case, we find a nonlinear ODE for $y(z)$. Let us try to find a solution in the simpler case that $Q(z)=Q_0$, this becomes:
\begin{equation}
y''(z) - \frac{2y'(z)}{1+z} = -\frac{8\pi G Q_0}{(1+z)^2 \sqrt{y(z)}}
\label{eqyQ0}
\end{equation}
Which yet is a nonlinear ODE for $y(z)$ and has not an analytical closed form solution. Let us, then, assume that the interaction $Q(z)=Q_0$ is small, and assume, at zeroth-order, an approximate solution of the form:
\begin{equation}
y(z) = H_0^2 [\Omega_\Lambda + \Omega_m (1+z)^3]
\end{equation}
where $\Omega_\Lambda + \Omega_m = 1$. Substituting this into the rhs of equation \eqref{eqyQ0} yields a first order $y_1$ solution with correction terms due to $Q_0$. The general solution takes the form:
\begin{align}
E^2&\simeq E_\Lambda(z)^2
+\frac{\mathcal{Q}_0 E_{\Lambda}(z)}{3\Omega_\Lambda}
-\frac{\mathcal{Q}_0 \Omega_m(1+z)^3}{3\Omega_\Lambda^{3/2}}\tanh^{-1}\left(\frac{E_{\Lambda}(z)}{\sqrt{\Omega_\Lambda}}\right)
-\frac{2\mathcal{Q}_0}{3\sqrt{\Omega_\Lambda}} \tanh^{-1}\left(\frac{E_{\Lambda}(z)}{\sqrt{\Omega_\Lambda}}\right)-\nonumber\\
&-\frac{(1+z)^3\mathcal{Q}_0}{3\Omega_\Lambda^{3/2}}\left[\sqrt{\Omega_\Lambda}
-\Omega_m \tanh^{-1}\left(\frac{1}{\sqrt{\Omega_\Lambda}}\right)\right]+\frac{2\mathcal{Q}_0}{3\sqrt{\Omega_\Lambda}}\,
   \tanh^{-1}\left(\frac{1}{\sqrt{\Omega_\Lambda}}\right)
   \label{E2aprox}
\end{align}
where $E\equiv\frac{H}{H_0}$ as before, $E_\Lambda(z) = \sqrt{\Omega_\Lambda + \Omega_m (1+z)^3}$ is the zeroth-order solution, $\Omega_\Lambda=1-\Omega_m$ and we have already replaced $y=H^2$ and the initial conditions:
\begin{align}
y(0) &= H_0^2 \\
y'(0) &= 3H_0^2 \Omega_m
\end{align}

\begin{figure}[htbp]
    \centering
    \includegraphics[width=0.49\textwidth]{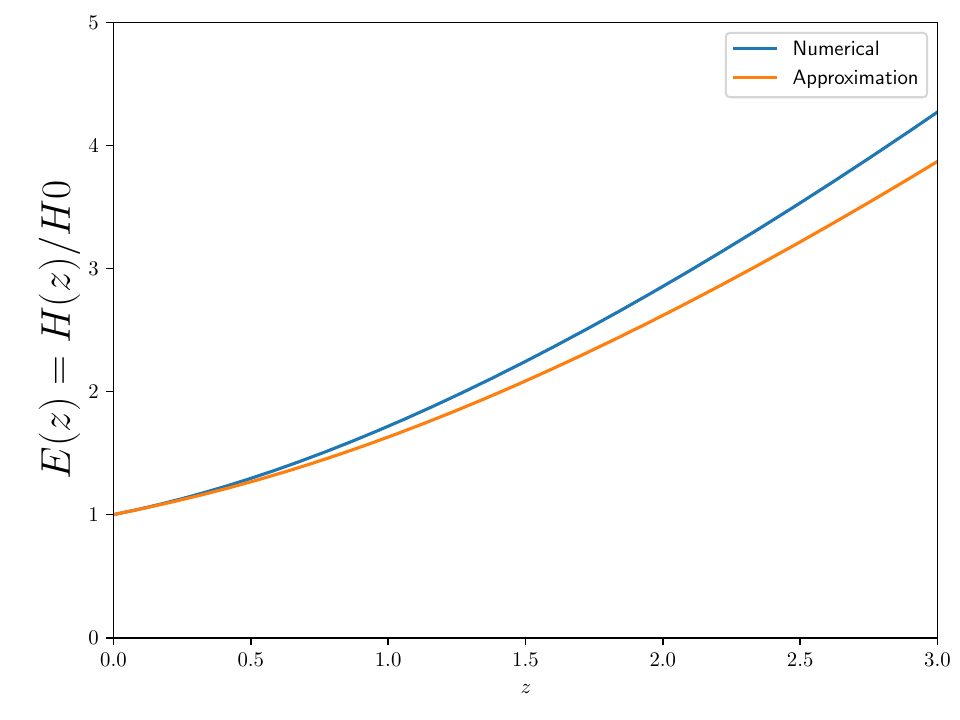}
    \includegraphics[width=0.49\textwidth]{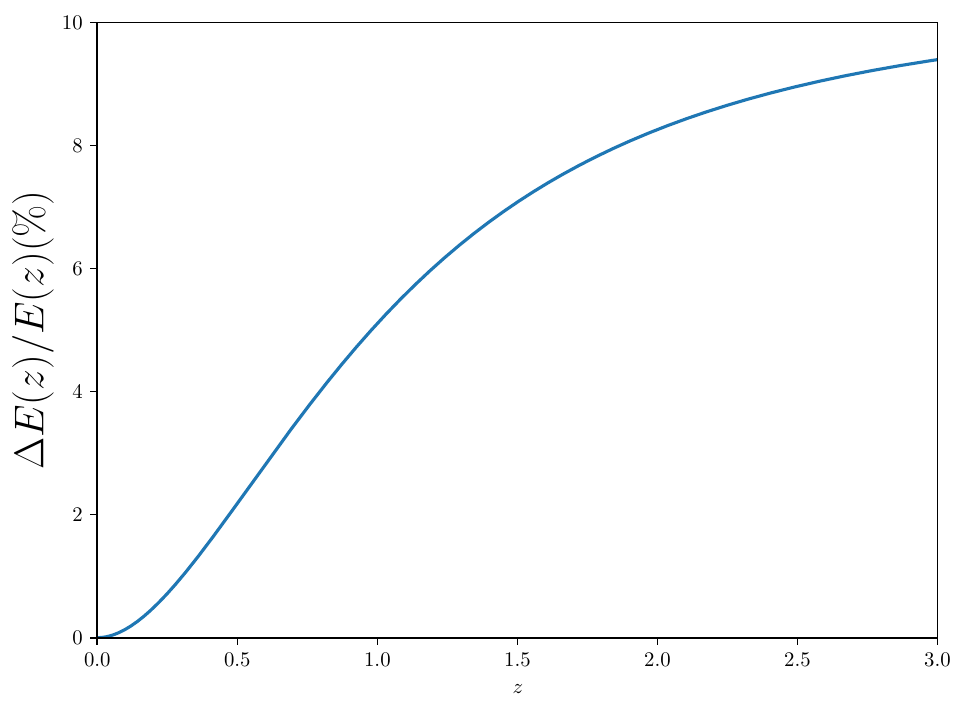}
    \caption{\textbf{Left:} Comparison of the normalized Hubble parameter $E(z) = H(z)/H_0$ obtained from the numerical and approximated solutions for the model $Q(z) = Q_0$. \textbf{Right:}  Relative deviation of $E(z)$ as a function of redshift for the model $Q(z) = Q_0$. The vertical axis shows the percentage difference $\Delta E(z)/E(z)$ between the numerical and approximated solutions.}
    \label{fig:com}
\end{figure}

In order to compare the approximation \eqref{E2aprox} with the exact numerical solution for $E(z)$ from Eqs. \eqref{Ezsys1}-\eqref{Ezsys2} with $Q(z)=Q_0$, we have plotted both results in Fig. \ref{fig:com}. In this figure, we have assumed the best fit parameters from Tab. \ref{tab:Q0}.

\newpage

As we can see from this Figure, the approximation for $E(z)$ essentially follows the numerical result, only resulting in a slightly lower value for $E(z)$ at high redshifts, but not deviating from the exact $E(z)$ for more than 10\%, in the redshift interval of the available data. We shall emphasize, however, that in the statistical analysis of Sec. \ref{sec: analysis}, we have used only the exact solution for $E(z)$, as the data allows for a large variation of $Q_0$, as can be seen on Tab. \ref{tab:Q0}.

\newpage
\begin{acknowledgments}
This study was financed in part by the Coordena\c{c}\~ao de Aperfei\c{c}oamento de Pessoal de N\'ivel Superior - Brasil (CAPES) - Finance Code 001. JFJ acknowledges financial support from  {Conselho Nacional de Desenvolvimento Cient\'ifico e Tecnol\'ogico} (CNPq) (No. 314028/2023-4). RV is supported by  Funda\c{c}\~ao de Amparo \`a Pesquisa do Estado de S\~ao Paulo - FAPESP (thematics projects process no. 2021/13757-9 and no. 2021/01089-1).
\end{acknowledgments}

\bibliographystyle{apsrev}
\bibliography{references}

@book{weinberg1972gravitation,
  title={Gravitation and Cosmology: Principles and Applications of the General Theory of Relativity},
  author={Weinberg, Steven},
  year={1972},
  publisher={John Wiley \& Sons}
}

@book{peebles1993principles,
  title={Principles of Physical Cosmology},
  author={Peebles, P. J. E.},
  year={1993},
  publisher={Princeton University Press}
}

@article{lima2025new,
    author = "Lima, P. W. R. and Lima, J. A. S. and Jesus, J. F.",
    title = "{New accelerating cosmology without dark energy: the particle creation approach and the reduced relativistic gas}",
    eprint = "2502.14139",
    archivePrefix = "arXiv",
    primaryClass = "astro-ph.CO",
    doi = "10.1140/epjc/s10052-025-14115-y",
    journal = "Eur. Phys. J. C",
    volume = "85",
    number = "4",
    pages = "449",
    year = "2025"
}

@article{jesus2025high,
  title={High-redshift cosmography with a possible cosmic distance duality relation violation},
  author={Jesus, Jos{\'e} F and Gomes, Mikael JS and Holanda, Rodrigo FL and Nunes, Rafael C},
  journal={Journal of Cosmology and Astroparticle Physics},
  volume={2025},
  number={01},
  pages={088},
  year={2025},
  publisher={IOP Publishing}
}

@article{StaicovaBenisty21,
    author = "Staicova, Denitsa and Benisty, David",
    title = "{Constraining the dark energy models using baryon acoustic oscillations: An approach independent of H0 \ensuremath{\cdot} rd}",
    eprint = "2107.14129",
    archivePrefix = "arXiv",
    primaryClass = "astro-ph.CO",
    doi = "10.1051/0004-6361/202244366",
    journal = "Astron. Astrophys.",
    volume = "668",
    pages = "A135",
    year = "2022"
}

@article{Pantheon+,
    author = "Brout, Dillon and others",
    title = "{The Pantheon+ Analysis: Cosmological Constraints}",
    eprint = "2202.04077",
    archivePrefix = "arXiv",
    primaryClass = "astro-ph.CO",
    doi = "10.3847/1538-4357/ac8e04",
    journal = "Astrophys. J.",
    volume = "938",
    number = "2",
    pages = "110",
    year = "2022"
}

@article{MorescoEtAl22,
    author = "Moresco, Michele and others",
    title = "{Unveiling the Universe with emerging cosmological probes}",
    eprint = "2201.07241",
    archivePrefix = "arXiv",
    primaryClass = "astro-ph.CO",
    doi = "10.1007/s41114-022-00040-z",
    journal = "Living Rev. Rel.",
    volume = "25",
    number = "1",
    pages = "6",
    year = "2022"
}

@article{emcee,
  title={emcee: the MCMC hammer},
  author={Foreman-Mackey, Daniel and Hogg, David W and Lang, Dustin and Goodman, Jonathan},
  journal={Publications of the Astronomical Society of the Pacific},
  volume={125},
  number={925},
  pages={306},
  year={2013},
  publisher={IOP Publishing}
}

@article{PressEtAl92,
  title={Numerical recipes in Fortran 77},
  author={Press, William H and Teukolsky, Saul A and Vetterling, William T and Flannery, Brian P},
  journal={The art of scientific computing},
  volume={1},
  year={1992}
}

@article{OverdCooper98,
    author = "Overduin, J. M. and Cooperstock, F. I.",
    title={Evolution of the scale factor with a variable cosmological term},
   volume={58},
   ISSN={1089-4918},
   DOI={10.1103/physrevd.58.043506},
   number={4},
   journal={Physical Review D},
   publisher={American Physical Society (APS)},
   year={1998},
   month=jul 
}

@phdthesis{teseJesus,
  title={Energia escura e acelera{\~A} {\S} {\~A}{\pounds} o do Universo: Aspectos conceituais e testes observacionais},
  author={Jesus, J. F.},
  year={2010},
  school={Universidade de S{\~a}o Paulo}
}

@article{Bronstein33,
  author       = {Bronstein, M.},
  title        = {Zur Frage der Gravitationstheorie},
  journal      = {Physikalische Zeitschrift der Sowjetunion},
  year         = {1933},
  volume       = {3},
  pages        = {73},
}

@article{OzerTaha86,
    author = "Ozer, Murat and Taha, M. O.",
    title = "{A Solution to the Main Cosmological Problems}",
    reportNumber = "Print-85-0485 (RIYADH)",
    doi = "10.1016/0370-2693(86)91421-8",
    journal = "Phys. Lett. B",
    volume = "171",
    pages = "363--365",
    year = "1986"
}

@article{ChenWu90,
    author = "Chen, W. and Wu, Y. S.",
    title = "{Implications of a cosmological constant varying as R**(-2)}",
    doi = "10.1103/PhysRevD.41.695",
    journal = "Phys. Rev. D",
    volume = "41",
    pages = "695--698",
    year = "1990",
    note = "[Erratum: Phys.Rev.D 45, 4728 (1992)]"
}

@article{CarvalhoEtAl92,
    author = "Carvalho, J. C. and Lima, J. A. S. and Waga, I.",
    title = "{On the cosmological consequences of a time dependent lambda term}",
    reportNumber = "IF-UFRJ-91-36",
    doi = "10.1103/PhysRevD.46.2404",
    journal = "Phys. Rev. D",
    volume = "46",
    pages = "2404--2407",
    year = "1992"
}

@article{Abdel-Rahman92,
    author = "Abdel-Rahman, A. M. M.",
    title = "{Singularity - free decaying vacuum cosmologies}",
    doi = "10.1103/PhysRevD.45.3497",
    journal = "Phys. Rev. D",
    volume = "45",
    pages = "3497--3511",
    year = "1992"
}

@article{Jesus06,
    author = "Jesus, Jose Fernando",
    title = "{Constraints from the old quasar apm 08279+5255 on two classes of lambda(t)-cosmologies}",
    eprint = "astro-ph/0603142",
    archivePrefix = "arXiv",
    doi = "10.1007/s10714-007-0533-0",
    journal = "Gen. Rel. Grav.",
    volume = "40",
    pages = "791--798",
    year = "2008"
}

@article{Rajeev83,
    author = "Rajeev, S. G.",
    title = "{Why Is the Cosmological Constant Small?}",
    reportNumber = "SU-4217-246, COO-3533-246",
    doi = "10.1016/0370-2693(83)91255-8",
    journal = "Phys. Lett. B",
    volume = "125",
    pages = "144--146",
    year = "1983"
}

@article{Riess22,
    author = "Riess, Adam G. and others",
    title = "{A Comprehensive Measurement of the Local Value of the Hubble Constant with 1 km s$^{−1}$ Mpc$^{−1}$ Uncertainty from the Hubble Space Telescope and the SH0ES Team}",
    eprint = "2112.04510",
    archivePrefix = "arXiv",
    primaryClass = "astro-ph.CO",
    doi = "10.3847/2041-8213/ac5c5b",
    journal = "Astrophys. J. Lett.",
    volume = "934",
    number = "1",
    pages = "L7",
    year = "2022"
}

@article{Planck:2018vyg,
    author = "Aghanim, N. and others",
    collaboration = "Planck",
    title = "{Planck 2018 results. VI. Cosmological parameters}",
    eprint = "1807.06209",
    archivePrefix = "arXiv",
    primaryClass = "astro-ph.CO",
    doi = "10.1051/0004-6361/201833910",
    journal = "Astron. Astrophys.",
    volume = "641",
    pages = "A6",
    year = "2020",
    note = "[Erratum: Astron.Astrophys. 652, C4 (2021)]"
}

@article{DESI24,
    author = "Adame, A. G. and others",
    collaboration = "DESI",
    title = "{DESI 2024 VII: cosmological constraints from the full-shape modeling of clustering measurements}",
    eprint = "2411.12022",
    archivePrefix = "arXiv",
    primaryClass = "astro-ph.CO",
    reportNumber = "FERMILAB-PUB-24-0854-PPD",
    doi = "10.1088/1475-7516/2025/07/028",
    journal = "JCAP",
    volume = "07",
    pages = "028",
    year = "2025"
}

@article{dr623,
  author        = {Madhavacheril, Mathew S. and others},
  collaboration = {Atacama Cosmology Telescope},
  title         = {The Atacama Cosmology Telescope: DR6 Gravitational Lensing Map and Cosmological Parameters},
  journal       = {Astrophys. J.},
  volume        = {962},
  number        = {2},
  pages         = {113},
  year          = {2024},
  doi           = {10.3847/1538-4357/ad1ddd},
  eprint        = {2304.05203},
  archivePrefix = {arXiv},
  primaryClass  = {astro-ph.CO}
}

@article{DESI24s8,
    author = "Sailer, Noah and others",
    title = "{Cosmological constraints from the cross-correlation of DESI Luminous Red Galaxies with CMB lensing from Planck PR4 and ACT DR6}",
    eprint = "2407.04607",
    archivePrefix = "arXiv",
    primaryClass = "astro-ph.CO",
    doi = "10.1088/1475-7516/2025/06/008",
    journal = "JCAP",
    volume = "06",
    pages = "008",
    year = "2025"
}

@article{YangEtAl15,
  author       = {Yang, Tao and Guo, Zong-Kuan and Cai, Rong-Gen},
  title        = {Reconstructing the interaction between dark energy and dark matter using Gaussian Processes},
  journal      = {Physical Review D},
  volume       = {91},
  number       = {12},
  pages        = {123533},
  year         = {2015},
  doi          = {10.1103/PhysRevD.91.123533},
  eprint       = {1505.04443},
  archivePrefix= {arXiv},
  primaryClass = {astro-ph.CO}
}

@article{MarttensEtAl20,
  author       = {von Marttens, Rodrigo and Gonzalez, Javier E. and Alcaniz, Jailson and Marra, Valerio and Casarini, Luciano and others},
  title        = {A model-independent reconstruction of dark sector interactions},
  journal      = {Physical Review D},
  volume       = {103},
  number       = {10},
  pages        = {103531},
  year         = {2021},
  doi          = {10.1103/PhysRevD.103.103531},
  eprint       = {2011.10846},
  archivePrefix= {arXiv},
  primaryClass = {astro-ph.CO}
}

@article{BonillaEtAl22,
  author       = {Bonilla, Alexander and Cárdenas, Victor H. and Motta, Verónica},
  title        = {Reconstruction of the dark sectors' interaction: A model-independent inference and forecast from GW standard sirens},
  journal      = {European Physical Journal C},
  volume       = {82},
  number       = {12},
  pages        = {1135},
  year         = {2022},
  doi          = {10.1140/epjc/s10052-022-11021-2},
  eprint       = {2102.06149},
  archivePrefix= {arXiv},
  primaryClass = {astro-ph.CO}
}

@article{EscamillaEtAl23,
  author       = {Escamilla, Luis A. and Akarsu, {\"O}zg{\"u}r and Di Valentino, Eleonora and V{\'a}zquez, J. Alberto},
  title        = {Model-independent reconstruction of the interacting dark energy kernel: Binned and Gaussian process},
  journal      = {Journal of Cosmology and Astroparticle Physics},
  volume       = {2023},
  number       = {11},
  pages        = {006},
  year         = {2023},
  doi          = {10.1088/1475-7516/2023/11/006},
  eprint       = {2307.03134},
  archivePrefix= {arXiv},
  primaryClass = {astro-ph.CO}
}

@article{Seikel:2012uu,
    author = "Seikel, Marina and Clarkson, Chris and Smith, Mathew",
    title = "{Reconstruction of dark energy and expansion dynamics using Gaussian processes}",
    eprint = "1204.2832",
    archivePrefix = "arXiv",
    primaryClass = "astro-ph.CO",
    doi = "10.1088/1475-7516/2012/06/036",
    journal = "JCAP",
    volume = "06",
    pages = "036",
    year = "2012"
}

\end{document}